\documentclass[pra,twocolumn,floatfix,superscriptaddress]{revtex4-1}
\usepackage{dcolumn,amsmath}
\usepackage{graphicx}
\usepackage{bm}
\usepackage{amssymb}
\usepackage{multirow}

\usepackage{amsfonts}
\usepackage{amsmath}
\usepackage{cancel}
\usepackage{xcolor}
\begin{document}
\title{Electric field dependent g factors of RaOCH$_3$ molecule}
\author {Alexander Petrov}\email{petrov\_an@pnpi.nrcki.ru}
\affiliation{Petersburg Nuclear Physics Institute named by B.P. Konstantinov of National Research Centre
"Kurchatov Institute", Gatchina, 1, mkr. Orlova roshcha, 188300, Russia}
\affiliation{St. Petersburg State University, St. Petersburg, 7/9 Universitetskaya nab., 199034, Russia} 

\date{Received: date / Revised version: date}

\begin{abstract}{
The sensitivity of experiments searching for the electron electric dipole moment (eEDM) using the symmetric top molecules can be greatly enhanced by laser cooling.
A detailed understanding of the Zeeman structure of the eEDM-sensitive levels is crucial for controlling systematic effects.
 We have developed a method for calculating the $g$-factors of $K$-doublet levels in symmetric top molecules and applied it to RaOCH$_3$. The electric-field-dependent $g$-factors of the first excited rotational level of RaOCH$_3$ are calculated. $K$-doublet levels with a small difference in $g$-factors are identified, and the main contributions to this difference are determined.
}
\end{abstract}
\maketitle
\section{Introduction}
Measuring the electron electric dipole moment ($e$EDM) is now considered one of the most promising tests for physics beyond the Standard Model \cite{Fukuyama2012,PospelovRitz2014,YamaguchiYamanaka2020,YamaguchiYamanaka2021}.
In $e$EDM experiments using molecules, the energy splitting between levels with opposite projections of the total angular momentum is measured in the presence of electric and magnetic fields that are either parallel or antiparallel to each other. 

In addition to the $e$EDM contribution, there is a Zeeman-induced contribution to the splitting. Thus, insufficient control of the magnetic field is a source of systematic errors in the experiment.
Diatomic molecules with $\Omega$-doubling structure \cite{DeMille2001}, linear triatomic molecules in the first excited bending vibrational mode with $l$-doubling structure \cite{Kozyryev:17}, and symmetric top molecules with $K$-doubling structure \cite{Kozyryev:17} are highly robust against these systematics, since the $\Omega$-, $l$-, and $K$-doublet structures are arranged such that the $e$EDM contribution to the splitting has opposite signs in the two doublet states, whereas the Zeeman contributions have the same sign. Thus, the energy splittings in the two doublet states can be subtracted from each other, which suppresses systematic effects related to stray magnetic fields (and many other systematic effects) while doubling the $e$EDM signal.
As a consequence, significant progress in $e$EDM searches using HfF$^+$ \cite{newlimit1} and ThO \cite{ACME:18} is closely related to the $\Omega$-doubling structure of these molecules.
Unfortunately, the diatomic molecules considered here feature a complex electronic structure due to $\Omega$-doubling. In contrast, linear triatomic and symmetric top molecules in the ground electronic state are amenable to laser cooling, which can further improve the statistical sensitivity of the experiment by up to three orders of magnitude \cite{Isaev:16, Isaev_2017}.

Although the two components of the doublet structures are very similar, they have slightly different magnetic $g$-factors. Systematic effects related to magnetic field imperfections are suppressed by a factor of $\sim \Delta g / g$. The difference $\Delta g$ depends on the laboratory electric field and is an essential parameter for experimental planning.

Previously, the dependence of $g$-factors on the electric field was studied for many diatomic molecules, including PbO \cite{Petrov:11}, WC \cite{Lee:13a}, ThO \cite{Petrov:14}, HfF$^+$ \cite{Petrov:17b, Kurchavov:2020, Kurchavov2021}, RaF \cite{Petrov:2020}, and PbF \cite{Baturo:2021}. Recently, a method was developed and the $g$-factor difference between the two components of the $l$-doublet was calculated for the YbOH molecule \cite{Petrov:24b}.

Great progress has already been achieved in both theoretical and experimental studies of symmetric top molecules \cite{Zhang:21, zakharova22, Augenbraun:21, Gaul:24,  Yu:2021, Fan:25,  Petrov:2025}. However, to the best of our knowledge, there are currently no data on the $g$-factor difference between the two components of the $K$-doublet for symmetric top molecules. There is still no clarity regarding either the order of magnitude of $\Delta g$ or the specific contributions to the difference. In the present work, we have developed the method and calculated the $g$-factors of the first excited rotational level of RaOCH$_3$.

\section{Electron electric dipole moment sensitive levels}
The focus of the $e$EDM experiment is the $N=1$ rotational level, for which the quantum number $K$ takes the values $-1, 0, 1$.  $K$ is the projection of the rotational momentum on the symmetry axis. Hence, the $N=1$ rotational level is the lowest one exhibiting a $K$‑doubling structure.
We consider the RaOCH$_3$ molecule with spinless isotopes of Ra, O, and C, whereas the hydrogen atoms possess non‑zero nuclear spins $i = 1/2$. The total spin of the hydrogen nuclei is either $I = 1/2$ (two states) or $I = 3/2$ (one state).
Owing to the Pauli principle for the hydrogen nuclei, there is a strong correlation between the $K$ and $I$ quantum numbers.
$K = 0$ ($|K|$ a multiple of 3 in general) corresponds to $I = 3/2$, while $|K| = 1$ ($|K|$ not a multiple of 3 in general) corresponds to $I = 1/2$. The $K$‑doublet levels of definite parity are given by
\begin{equation}
|K=+1, 1I=1/2\rangle \pm |K=-1, 2I=1/2\rangle,
\end{equation}
where $|1I=1/2\rangle$ and $|2I=1/2\rangle$ are the hydrogen nuclear spin wavefunctions. Owing to the orthogonality of $|1I=1/2\rangle$ and $|2I=1/2\rangle$, only the inclusion of the hyperfine interaction can give rise to the $K$‑doubling effect \cite{Petrov:2025}.

The spin-rotation interaction of the electron spin ($S=1/2$) with the rotational momentum ($N=1$) gives rise to splitting between the energy levels with total electronic-rotational momenta $J = 1/2$ and $J = 3/2$. Finally, $J = 1/2$ and $J = 3/2$ couple with $I=1/2$ \footnote{We do not consider $I=3/2$ here, since these states do not form $K$-doublets for $N=1$, although the hyperfine interaction with $I=3/2$ states allowed by the Pauli principle is taken into account in the calculations. This has a negligible effect, however, because levels with different values of the $|K|$ quantum number have a large energy difference proportional to the rotational constant $A$.} to yield $|J = 1/2, F=0,1\rangle$ and $|J = 3/2, F=1,2\rangle$.

Levels with $M_F = 0$ ($M_F$ is the projection of the total momentum onto the laboratory axis) are insensitive to the $e$EDM. Thus, the focus of the present work will be on two $M_F = 2$ levels (one $K$-doublet) and six $M_F = 1$ levels (three $K$-doublets). According to the estimates in Ref. \cite{Petrov:2025}, the $K$-doubling values are 13.6 kHz for $|N=1, J=3/2, F=1\rangle$, 8.2 kHz for $|N=1, J=3/2, F=2\rangle$, and 26.1 kHz for $|N=1, J=1/2, F=1\rangle$. Owing to the extreme closeness of these levels, they are almost completely polarized by a very small electric field $E \approx 300$ mV/cm.

\section{Method}

Following Ref.~\cite{Petrov:2025}, we present our Hamiltonian in the molecular reference frame as
\begin{equation}
{\rm \bf\hat{H}} = {\rm \bf\hat{H}}_{\rm mol} + {\rm \bf\hat{H}}_{\rm hfs} + {\rm \bf\hat{H}}_{\rm S} + {\rm \bf\hat{H}}_{\rm Z},
\label{Hamtot}
\end{equation} 
where
\begin{equation}
\hat{\rm H}_{\rm mol}= B_{\rm r}(\hat{\bf J} -\hat{\bf J}^{e} )^2+ \left(A-B_{\rm r}\right)(\hat{J}_z -\hat{ J}^{e}_z)^2
\label{Hmolf}
\end{equation}
is the molecular Hamiltonian. $B_{\rm r}= 0.0631$ cm$^{-1}$ and $A=4.83$ cm$^{-1}$ are rotational constants corresponding to the equilibrium geometry obtained in Ref.~\cite{zakharova22}.
$\hat{\bf J}^{e}$ is the electronic momentum.

${\rm \bf\hat{H}}_{\rm hfs}$ and ${\rm \bf\hat{H}}_{\rm S}$ are the hyperfine interaction with the H nuclei and the Stark interaction with the external electric field, respectively. In the current work, we have fixed the parameters of ${\rm \bf\hat{H}}_{\rm mol}$, ${\rm \bf\hat{H}}_{\rm hfs}$ and ${\rm \bf\hat{H}}_{\rm S}$ from Ref.~\cite{Petrov:2025}.

\begin{equation}
 {\rm \bf\hat{H}}_{\rm Z} = \mu_{\rm B}\hat{ {\bf G}}\cdot{\bf B} -\sum_{k=1}^{3}{\rm g}_{\rm H}\mu_{N}{\bf \hat{i}^{\rm k}}\cdot{\bf B}
 \label{HZe}
\end{equation}
describes the interaction of the molecule with an external magnetic field ${\bf B}$. Here, $\hat{ {\bf G}} = \hat{ {\bf L}}-{\rm g}_{S}\hat{ {\bf S}}$, $\hat{ \bf L}$ is the electronic orbital angular momentum operator, ${\rm g}_{S} = -2.0023$ is the free-electron $g$-factor, and $g_{\rm H} = 2.7928456$ is the $g$-factor of the hydrogen nucleus, $\mu_{B}$ and $\mu_{N}$ are the Bohr and nuclear magnetons, respectively.

Wavefunctions, energies, and $g$-factors were obtained by numerical diagonalization of the Hamiltonian (\ref{Hamtot}) over the basis set of electronic-rotational-vibrational-nuclear spin wavefunctions. We start from the basis set
\begin{equation}
 \Psi^e_{\Omega }\Theta^{J}_{M_J,\omega}(\alpha,\beta,\gamma)U^{1}_{M^{1}_i}U^{2}_{M^{2}_i}U^{3}_{M^{3}_i},
\label{basis}
\end{equation}
where
 $\Theta^{J}_{M_J,\omega}(\alpha,\beta, \gamma)=\sqrt{(2J+1)/{8\pi^2}}D^{J}_{M_J,\omega}(\alpha,\beta,\gamma)$ is the rotational wavefunction, $\alpha,\beta, \gamma$ are Euler angles defining the orientation of the molecular frame $xyz$ relative to the laboratory frame $XYZ$. The $z$ axis corresponds to the symmetry axis of RaOCH$_3$, and the $x$ axis is oriented such that the first hydrogen atom lies in the $xz$ plane, $U^{k}_{M^{k}_i}$ are the hydrogen nuclear spin wavefunctions, $M_J$ is the projection of the molecular (electronic-rotational) angular momentum $\hat{\bf J}$ onto the $Z$ axis, $\omega$ is the projection of the same momentum onto the $z$ axis, and $M^{k}_i$ are the projections of the hydrogen nuclear angular momenta onto the $Z$ axis, $\Psi^e_{\Omega}$ is the electronic wavefunction, and $\Omega \approx \pm 1/2$ is the projection of the electronic angular momentum onto the $z$ axis. Then, following Ref.~\cite{Petrov:2025} and using the approximate relation $K=\omega-\Omega$, we construct proper antisymmetric basis functions that satisfy the Pauli principle for $N=0, 1, 2$.

In the molecular frame coordinate system, the molecular (\ref{Hmolf}) and Zeeman (\ref{HZe}) interactions are determined by the electronic matrix elements $\hat{J}^{e}_z$, $\hat{J}^{e}_+ = \hat{J}^{e}_x + i\hat{J}^{e}_y$, and $\hat{G}_z$, $\hat{G}_+ = \hat{G}_x + i\hat{G}_y$, respectively. Their matrices (where the first basis function is $\Psi^e_{\Omega=-1/2}$ and the second is $\Psi^e_{\Omega=1/2}$) are

\begin{equation}
\hat{J}^{e}_z = 
\begin{pmatrix}
-0.49986 &  0.0000 \\
0.0000 &  0.49986
\end{pmatrix},
\label{Jz}
\end{equation}

\begin{equation}
\hat{J}^{e}_+ = 
\begin{pmatrix}
0.0000 &  0.0000 \\
 1.0342 & 0.0000
\end{pmatrix},
\label{Jp}
\end{equation}

\begin{equation}
\hat{G}_z = 
\begin{pmatrix}
-0.99672 &  0.0000 \\
0.0000 & 0.99672
\end{pmatrix},
\label{Gz}
\end{equation}

\begin{equation}
\hat{G}_+ = 
\begin{pmatrix}
0.0000 &  0.0000 \\
 2.0321 & 0.0000
\end{pmatrix}.
\label{Gp}
\end{equation}
Matrices (\ref{Jz}) and (\ref{Jp}) were taken from Ref.~\cite{Petrov:2025}. Matrices (\ref{Gz}) and (\ref{Gp}) were calculated on the basis of matrix elements of $\hat{ {\bf L}}$ and $\hat{ {\bf S}}$ computed within the same approximation.

\section{Results}
We define the effective $g$-factors such that Zeeman shift is equal to
\begin{equation} 
   E_{\rm Z} = {\rm g}\mu_B B M_F.
 \label{Zeem}
\end{equation}

Calculated values for the average $g$-factors, $(g_u + g_l)/2$, and their difference, $\Delta g = g_u - g_l$, of $K$-doublets are $0.84$ and $4.7\cdot10^{-4}$ for $N=1,J=3/2,F=1$; $0.50$ and $-1.5\cdot10^{-8}$ for $N=1,J=3/2,F=2$; $-0.33$ and $-4.7\cdot10^{-4}$ for $N=1,J=1/2,F=1$. We use the designation $g_u$ ($g_l$) for the $g$-factors of the upper (lower) level of the $K$-doublet.

Let us analyze the perturbations leading to the difference in $g$-factors.
The difference $\Delta g$ has the same magnitude and opposite sign for $N=1,J=3/2,F=1$ and $N=1,J=1/2,F=1$, which indicates that these states perturb each other via hyperfine interaction. The matrix element of the hyperfine interaction between $N=1,J=3/2,F=1$ and $N=1,J=1/2,F=1$, though formally allowed, is equal to zero for the exact $N, S$ quantum numbers. Indeed, let us consider another coupling scheme where the electron spin ($S=1/2$) and the hydrogen nuclear spin ($I=1/2$) couple to the angular momentum $G=0,1$; then $G=0$ and $N=1$ couple to the total momentum $F=1$ whereas $G=1$ and $N=1$ couple to the total momenta $F=0,1,2$. We see that the two $F=1$ states have different quantum numbers $G$ and do not interact. The exact $N, S$ quantum numbers correspond to matrix elements (compare with (\ref{Jz}) and (\ref{Jp}))
\begin{equation}
\hat{J}^{e}_z =
\begin{pmatrix}
-0.5 & 0.0 \\
0.0 & 0.5
\end{pmatrix},
\label{Jz0}
\end{equation}

\begin{equation}
\hat{J}^{e}_+ =
\begin{pmatrix}
0 & 0 \\
1 & 0
\end{pmatrix}.
\label{Jp0}
\end{equation}
Our calculations confirm that in this case $\Delta g$ is about three orders of magnitude smaller.
Thus, taking into account the molecular interaction (\ref{Hmolf}) between $|N=1,J=3/2\rangle$ and $|N=2,J=3/2\rangle$ is very important for the accurate evaluation of $\Delta g$.

For the calculation of the matrix element between these states it is convenient to write the states in Hund's case (c) coupling scheme \cite{Petrov:2025}:
\begin{eqnarray}
\nonumber
| K=1,N=1,S,J=3/2,M_J\rangle = \\
\nonumber
\sqrt{3/4},|\Omega=1/2,K=1,J=3/2,M_J\rangle + \\
\nonumber
\sqrt{1/4},|\Omega=-1/2,K=1,J=3/2,M_J\rangle,
\label{WF32}
\end{eqnarray}
\begin{eqnarray}
\nonumber
| K=1,N=2,S,J=3/2,M_J\rangle = \\
\nonumber
-\sqrt{1/4},|\Omega=1/2,K=1,J=3/2,M_J\rangle + \\
\nonumber
\sqrt{3/4},|\Omega=-1/2,K=1,J=3/2,M_J\rangle.
\label{WF32b}
\end{eqnarray}

Then one can calculate that
\begin{eqnarray}
\nonumber
\langle N=1,J=3/2| \hat{\bf H}_{\text{mol}} |N=2,J=3/2\rangle = \\
\sqrt{3}\left[A\left((\hat{J}^{e}_z)_{22} - 1/2\right) +
B_{\rm r}\left(1/2 - (\hat{J}^{e}_+)_{12}/2\right)\right].
\label{PT}
\end{eqnarray}
Although $A$ is many times larger than $B_{\rm r}$, the deviation of $(\hat{J}^{e}_+)_{12}/2$ from one half is about the same number of times larger than the deviation of $(\hat{J}^{e}_z)_{22}$. Therefore, both terms in square brackets of Eq.~(\ref{PT}) give comparable contributions. On the other hand, the deviation of $(\hat{J}^{e}_z)_{22}$ from one half is only about $0.03\%$, and this value is very difficult to predict accurately from \textit{ab initio} calculations. Therefore, our values for the $g$-factor difference should be considered as an order-of-magnitude estimate.

The difference in the $g$-factors for the $F=2$ state is about four orders of magnitude smaller than that for $F=1$. We found that only two small perturbations lead to this difference. First, the perturbation of $|N=1,J=3/2,F=2\rangle$ by $|N=2,J=3/2,F=2\rangle$ is slightly different for positive and negative parity states due to the small denominator difference caused by $K$-doubling effects. Second, $|N=2,J=3/2,F=2\rangle$ (the perturbing states) are in turn perturbed by $|N=2,J=5/2,F=2\rangle$ due to hyperfine interaction. This perturbation also differs for different parity states because of the $K$-doubling effect.

 We note that the smallness of the difference between the $g$-factors is due, among other factors, to the Pauli principle, which forbids some possible perturbing states. For example, for diatomics, the main effect in the field-free case is due to the admixture of $|\Omega=0 \rangle$ states to the $\Omega$-doublet because of the Coriolis interaction. Thus, the parity eigenfunctions of the $\Omega$-doublet components read $1 / \sqrt{2}(|\Omega=+1\rangle + |\Omega=-1\rangle) + \epsilon |\Omega=0^+\rangle$ for one component and $1 / \sqrt{2}(|\Omega=+1\rangle - |\Omega=-1\rangle) + {\epsilon}' |\Omega=0^-\rangle$ for the other, where $\epsilon \ll 1$ and $\epsilon' \ll 1$.
The perturbing states $|\Omega=0^+\rangle$ and $|\Omega=0^-\rangle$ are different electronic states for the different parity components of the $\Omega$-doublet. A similar effect also exists for triatomic molecules \cite{Petrov:24b}, though the projection of the vibrational momentum $m$ should be used instead of the projection of the electronic momentum $\Omega$. In the considered case, the molecular interaction (\ref{Hmolf}) does not mix the components $|K=+1, 1I=1/2\rangle \pm |K=-1, 2I=1/2\rangle$ of the $K$-doublet with $|K=0, I=3/2\rangle$ due to the orthogonality of the nuclear spin wavefunctions. This perturbation would, however, take place for the RaOCHDT molecule with different hydrogen isotopes.

\begin{figure}
\includegraphics[width=0.95\linewidth]{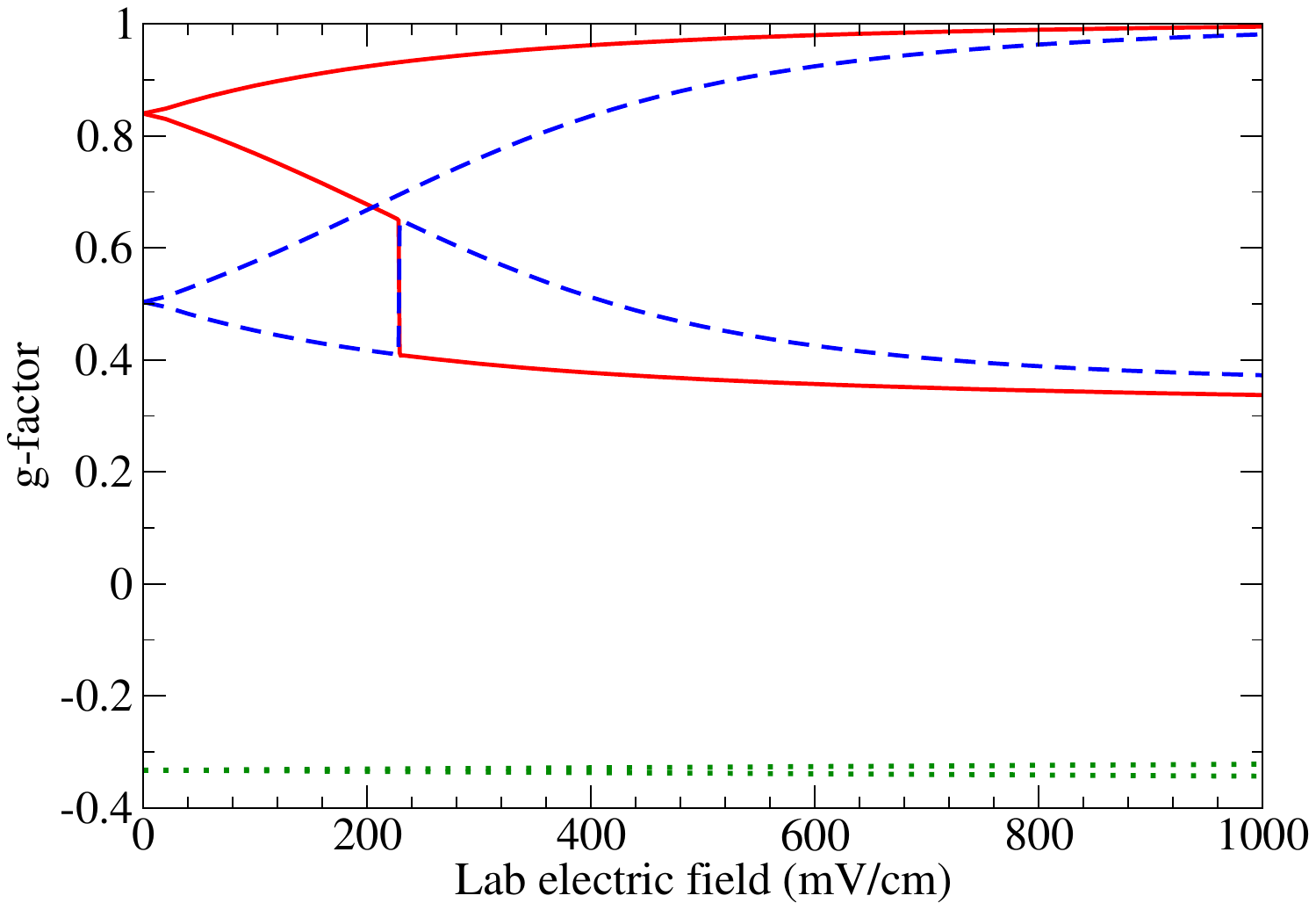}
 \caption{(Color online) Calculated $g$-factors 
 for the $N = 1, M_F=1$  Zeeman sublevels
 as functions of the laboratory electric field.}
 \label{gf1}
\end{figure}

\begin{figure}
\includegraphics[width=0.95\linewidth]{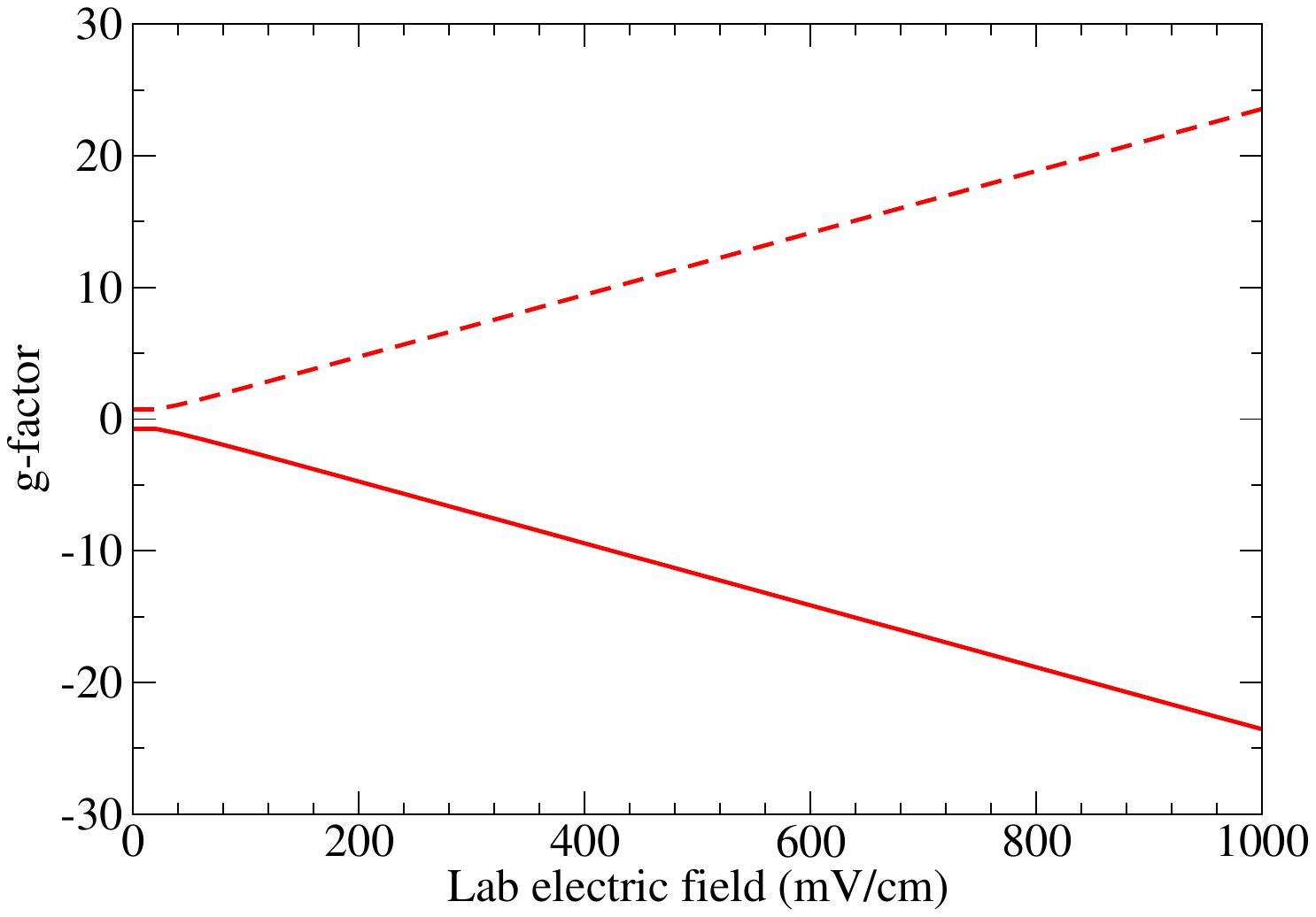}
 \caption{(Color online) Calculated $g$-factors 
 for the $N = 1, M_F=2$  Zeeman sublevels
 as functions of the laboratory electric field.
 The value $(g-0.50337)\cdot10^{8}$ is printed.
 The value $(g_u + g_l)/2 =.50337$ corresponds to the average value for the field free case.
 Solid line for upper $K$-doublet level, dashed line for the lower $K$-doublet level.}
 \label{gf2}
\end{figure}

In Figures \ref{gf1} and \ref{gf2}, the calculated $g$-factors for the $N = 1, M_F=1$ and $N = 1, M_F=2$ Zeeman sublevels, respectively, are shown as functions of the laboratory electric field $E < 1$ V/cm. It has been shown in Ref.~\cite{Petrov:2025} that already at an electric field $E \sim 250$ mV/cm, the polarization of the molecule is almost saturated. Thus, the electric field in the range from 250 mV/cm to 1 V/cm provides a maximal $e$EDM signal and is comfortable from an experimental point of view.
One can see that, in general, the differences in $g$-factors for the $M_F=1$ levels, due to complex interactions between them, become much larger than in the field-free case.
For $M_F=2$, the perturbing state is another rotational level; therefore, the difference in $g$-factors is much smaller compared to the $M_F=1$ levels. This difference increases almost linearly with the electric field. The same perturbation and similar behavior exist for diatomic \cite{Petrov:11} and triatomic molecules \cite{Petrov:2025}. Since the electric field required to polarize the molecule is small, the value of the $g$-factor difference is also small. For $E = 500$ mV/cm, it is $\Delta g = g_u - g_l \approx -2.3\times10^{-7}$. This value, with good accuracy, is independent of the initial even smaller difference (Due to electric-field mixing of parity components in the $K$-doublet, the small initial difference averages to zero \cite{Petrov:11}) and is much more stable in calculations, since the final small value (in contrast to the field-free case) is not a result of cancellations of large numbers.

The obtained ratio $\Delta { g}/{ g} \sim 10^{-6}$ can be compared to $\Delta { g}/{ g} \le 4 \cdot 10^{-5}$ for $^{174}$YbOH \cite{Petrov:24b} and $\Delta { g}/{ g} \sim 10^{-3}$ for ThO \cite{Petrov:14} and HfF$^+$ \cite{Petrov:17b}, and thus appears more favorable.

\section{Conclusion}
We determined the main contributions to the $g$-factor difference of $K$-doublet components in the first excited $N=1$ rotational level of RaOCH$_3$. We found that the ratio $\Delta { g}/{ g}$ for $K$-doublet components in the $N = 1$, $J=3/2$, $M_F=2$ Zeeman sublevels is $\sim 10^{-6}$ for an electric field in the range from 500 mV/cm to 1 V/cm.

\section{Acknowledgements}
Calculations of the $g$-factor difference in RaOCH$_3$ are supported by the Russian Science Foundation grant no. 24-12-00092.
%
%

\end{document}